\documentclass[letterpaper]{sigcomm-alternate}
\usepackage{epsfig,url}
\usepackage{amssymb}
\usepackage{amsmath}
\usepackage{amsfonts}
\usepackage{graphicx}
\usepackage{algorithmic}
\usepackage{color}
\usepackage{listing}
\usepackage{float}
\usepackage{amssymb}
\usepackage{epsfig}
\usepackage{subfigure}
\usepackage{color}
\usepackage{hyperref}
\usepackage{url}
\long\def\comment#1{}

\definecolor{red}{rgb}{1,0,0}

\title{From Rough Consensus To Automated Reasoning}

\author{
Zied Ben Houidi \\
Alcatel-Lucent Bell Labs, France \\
\email{zied.ben\_houidi@alcatel-lucent.com}
}

\begin{document}
\maketitle
\begin{abstract}
Unwritten languages today often have no official grammar, and are rather governed by ``unspoken rules''. Similarly, we argue that the young discipline of networking is still a practice that lacks a deep understanding of the rules that govern it. 
This situation results in a tremendous loss of time and efforts. First, since the rules are unspoken, they are not systematically reused. Second, since there is no grammar, it is impossible to assert if a sentence is correct. Comparing two networking approaches or solutions is sometimes a synonym of endless religious debates. Drawing the proper conclusion from this claim, we advocate that networking research should spend more efforts on better understanding its rules as a first step to automatically reuse them. To illustrate our claim, we focus in this paper on one specific networking problem, and show how different instances of this same problem were solved in parallel, resulting in different solutions, and no explicit knowledge reuse. 
\end{abstract}

 
\section{Networking is not a science}\label{Introduction}
The practice precedes historically the theory. The practice of music, for instance, appeared before the theory of music, and the practice of the language appeared before the grammar. Unwritten languages today still have no known grammar. Similarly, we argue that the practice of networking (today) precedes a yet to be constructed theory of networking (hopefully tomorrow). Just like any unwritten language which is more governed by "unspoken rules" that resulted from a slow evolutionary process, networking is still a young discipline that lacks a clear understanding of its axioms and the rules that govern it~\cite{networking-philosopher}.

This situation translates into a set of inconveniences in this discipline today. First, it is not straightforward for networking engineers or researchers to clearly define what problem does a given designed protocol solve exactly~\cite{what-bgp-solves, shenker-talk}. This is due to the evolutionary approximate ``trial and error'' way of reasoning about networking problems at large. Second, another drawback is that the same networking problem often has a plethora of solutions and standards that are very hard to compare. The huge number of competing IETF RFCs that tackle the same problem is a good example to illustrate this situation. The way to settle disagreements between these competing solutions is even more telling: ``Rough consensus and running code''~\cite{rough-concensus}. 

Unfortunately, these inconveniences result in a tremendous waste of time and efforts. Since the rules of thumb that govern the networking designer's thinking are ``unspoken'', they are not systematically reused; several iterations are therefore needed to converge to an acceptable solution. 

The networking community reacted to this problem in three ways. The first reaction was timid with very little work (See~\cite{networking-philosopher,Formal-networking,modeling-keshav} and the very few papers that cite them). It consisted in being aware of our lack of understanding of the core rules that govern our discipline and the lack of formal methods in general. The second reaction, which was more successful, consisted in simplifying networking as a practice. This was expressed by the Software Defined Networking movement and its will to simplify today's complex network operations by making the control of network elements as generic and as flexible as possible. Finally, the third reaction consisted in modeling only particular aspects of networking today. A notable example in this context is Griffin's work on metarouting and the modeling of the Internet's routing protocols~\cite{metarouting1}. 

This document is a position paper, backed up with technical arguments, that goes into the direction of the first reaction. We advocate that we should move away from this trial and error engineering approach of networking today towards a more systematic approach where the accumulated knowledge learned from over 40 years of experience with networking is automatically reused. We call for an ambitious yet rewarding long term research goals for networking: to replace, as much as possible, the brain of a networking expert by an artificially intelligent entity that can automatically reason about networking problems. 
Such an automated ``networking expert'' should receive as an input a problem definition expressed in a high level language, explores the design space and gives as an output a set of possible solutions. 
If SDN is about augmenting the network with flexible ``arms'' (networking elements) that can be easily controlled by software (written by humans), then what we call for is augmenting the network with an artificially intelligent brain that can reason about networking problems.

To make a case for our claim, this paper illustrates by example the potential of knowledge reuse in networking problems solving. In particular, we show how variations of the same networking problem, namely the VPN service provisioning, were solved in parallel, resulting in different solutions, and no explicit knowledge reuse. We then articulate a new research direction for networking. This new direction is arduous and too ambitious, but it may fundamentally change our discipline.


\section{The lack of knowledge reuse by example}\label{model}
In this paper, we take the example of provider provisioned VPNs to illustrate that there is clearly a room for knowledge reuse in reasoning about networking problems. In particular, we consider a simple model that we previously developed~\cite{zied-icnp} to rethink the VPN routing problem for a specific family of VPNs~\cite{rfc:4364}. 
We then show how the generic solution to this model fits other VPN technologies (1) for which it was not designed, and (2) that were developed separately, thus highlighting by example the potential of knowledge reuse.
The model we reuse in this paper is similar to the ``one big switch'' abstraction that was recently proposed in the SDN context~\cite{big_switch}. It sees a provider network as a big switch that takes traffic from one ingress port locator, processes it, and outputs it from one egress port location. 
We first briefly describe provider provisioned VPNs. 

\subsection{Provider Provisioned VPNs}\label{goal}
An enterprise network often has sites in distant locations and that need to interconnect. Instead of building its own interconnection infrastructure, an enterprise can rely on a Virtual Private Network (VPN) service provider that shares its infrastructure among different enterprises, thus reducing the connection costs for each single company. The main task of such a VPN service provider is hence to multiplex different networks on its infrastructure while ensuring isolation between the different multiplexed networks. 

\emph{Putting together the architecture, the protocols and their configuration in order to provide such a VPN service is a networking problem}. This problem turned out to be different from existing networking problems, which called for new solutions. It took few years, few IETF working groups (l1vpn,l2vpn,l3vpn), new protocols or extensions to existing ones as well as as augmenting network equipments with new primitives (e.g. per interface forwarding) to finally come up with a set of solutions for provider provisioned VPNs~\cite{rfc5251,rfc4664,rfc:4364,vpn-bgp,rfc4761,VR:draft,RT-const}. And since the design space is still not fully covered, new solutions to this problem are even still being developed today~\cite{MAC-VPN}.

There exists roughly three levels or ways to interconnect two networks. The first is to only supply a physical connection between them. The second is to connect them through a layer 2 equipment like a hub or a switch. The third is to interconnect them through an IP router. These three interconnection ``types'' define the three main families of provider provisioned VPNs. First, in layer 1 VPNs~\cite{rfc5251}, the VPN provider supplies or emulates a physical link (e.g. a circuit) between any pair of networks that need to interconnect. Second, in layer 2 VPNs~\cite{rfc4664}, the provider supplies a layer 2 connection: for the special case of Ethernet for example, the provider acts as a switch that interconnects the networks of the same enterprise. Finally, in layer 3 IP VPNs~\cite{VR:draft,rfc:4364}, the provider acts as a router to interconnect netwoks of the same enterprise. Roughly speaking, in respectively layer 3, 2 and 1 VPNs, the shared infrastructure acts as a router, a switch and a physical link.

\subsection{A simple provider network model}
\begin{figure}[tb]
	\centering
		\includegraphics[scale=0.3]{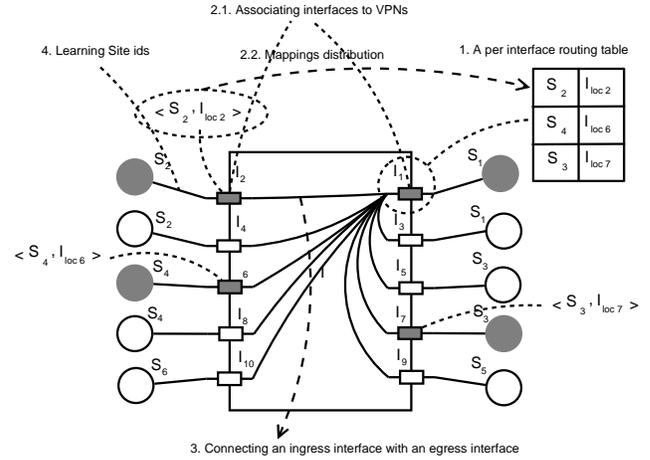}
	\caption{A simple provider network model}\label{fig:interface_matrix}
\end{figure}

We now briefly revisit the VPN provider network model that we previously developed~\cite{zied-icnp} to rethink BGP MPLS VPNs. 
The model, illustrated in Fig.~\ref{fig:interface_matrix}, represents the shared network by a set of provider \textit{interfaces} that connect VPN \textit{sites}. 
VPN interfaces are owned by the provider who has a full administrative control on them. Each provider interface connects a VPN site. A VPN is composed of a set of sites that need to be interconnected. A VPN site is identified by one (or more) \textit{site identifier(s)}, $S_{i}$. The site identifier addresses the VPN site and is local to each VPN (two sites pertaining to distinct VPNs can have the same identifier). The figure, illustrates an example of a provider network that connects two distinct VPNs, a white VPN and a gray VPN.
Each provider interface $I_{i}$ has an interface locator $I_{loc\; i}$. The locator locates the VPN interface inside the provider backbone. Indeed, since the provider is in control of his network, the model assumes that from any source interface, $I_{s}$, it is possible to send data to any destination interface $I_{d}$ given the only knowledge of its locator $I_{loc\; d}$.

\subsection{A generic VPN routing solution}\label{solution}
Based on this simple model, ~\cite{zied-icnp} sketched the set of elements that are necessary for the provider network to implement the IP VPN service. We summarize them as follows:

	\vspace{1mm}\noindent \textbf{1.~A per interface routing table}\footnote{Necessary because different VPNs can use the same site ids}: containing a \emph{mapping} between destination \emph{site ids} and corresponding egress \emph{interface locators}. Fig.~\ref{fig:interface_matrix} shows an example of the information that should be contained in the routing table of $I_{1}$. 
	
	\vspace{1mm}\noindent \textbf{2.~A method to populate per interface routing tables} with the correct mappings. This can be split in two tasks:
	
		\vspace{1mm}\noindent \textbf{2.1.~Associating interfaces to VPNs}: tagging similarly interfaces that are part of the same VPN in order to know locally (at each interface) which distant interfaces are part of the same VPN.
		
		\vspace{1mm}\noindent \textbf{2.2.~Mappings Distribution}: Sending locally known (site id,interface locator) mappings to be installed on similarly-tagged distant per-interface routing tables (of the same VPN).

In this paper, to be more exhaustive about the provider network needs, we add two other elements that were implicitly captured by the old model's assumptions and that are necessary for the VPN service provisioning:

	\vspace{1mm}\noindent \textbf{3.~Connecting an ingress interface with a specific egress interface} 
	
	\vspace{1mm}\noindent \textbf{4.~Learning which site ids are behind provider interfaces} 

Our thesis in this paper is that networking designers use ``unspoken rules'' to reason about networking problems, and that these rules are not reused while they could be. Solving or decomposing the VPN routing problem as we did on this simple model is an example of how a networking designer could reason about this networking problem. 
We next show that this ``effort'' could have been useful for other instances of the same problem. 
Indeed, although the model meant to rethink the particular case of BGP MPLS layer 3 IP VPNs, the effort it did fits other families of provider provisioned VPNs, that were solved separately. 
In particular, we show in the next sections that each of the other instances of the VPN routing problem exhibits a similar pattern: the four elements depicted above and curiously similar technical answers to address each of them.
We start with Layer 3 VPNs.

\subsection{The case of Layer-3 VPNs}
In layer-3 VPNs, a VPN site has a Customer Edge (CE) router that aggregates its traffic, and the provider network has a set of provider edge (PE) routers that connect the different enterprise CE routers. With respect to the model in Sec.~\ref{model}, the site id is one (or more) IP prefix(es) that address the VPN site network. Each PE router has a set of physical interfaces, each of them connects a VPN customer site.
These physical interfaces correspond to the interface in our reference model. We mainly find, in the literature, two approaches in layer-3 VPN solutions, the virtual router approach~\cite{VR:draft} and the BGP/MPLS IP VPNs one~\cite{rfc:4364}. They both have the same pattern as the generic solution depicted above in Sec.~\ref{solution}.

\subsubsection{BGP MPLS IP VPNs}\label{bgp_mpls}
\begin{figure}[h]
	\centering
		\includegraphics[scale=0.25]{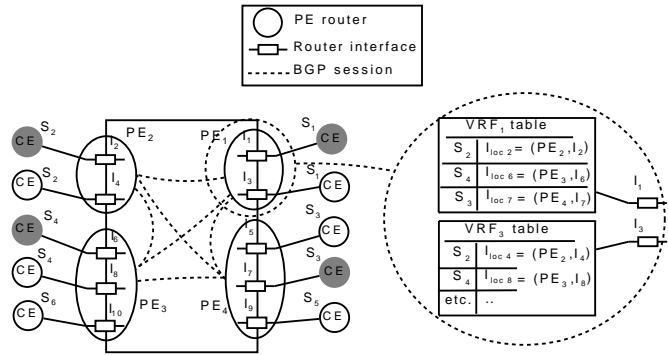}
	\caption{A BGP MPLS IP VPN implementation}\label{fig:Image9}
\end{figure}

Fig.~\ref{fig:Image9} illustrates the BGP/MPLS IP VPN approach to the VPN service on top of the provider network shown in Fig.~\ref{fig:interface_matrix}. This approach is built around the four elements depicted above:

	\vspace{1mm}\noindent \textbf{1.~A per interface routing table}: A PE associates by configuration a \emph{virtual routing and forwarding} (VRF) table per interface\footnote{In practice, the same VRF can be associated to more interfaces if they all belong to the same VPN}. The VRF table contains the mapping information (site id, interface locator) necessary for VPN routing. Fig.~\ref{fig:Image9} zooms on $PE_{1}$ and shows the content of $VRF_{1}$ and $VRF_{3}$ tables, corresponding respectively to $I_{1}$ and $I_{3}$. 
	Interface locators have two components: one that globally identifies the egress PE, and one that locally identifies the egress interface within the PE. This decomposition in two components is due to the way two interfaces are connected: we will see in point 3. below how this is realized. 
	
	\vspace{1mm}\noindent \textbf{2.~A method to populate per interface routing tables}: In BGP MPLS VPNs, this is done using multi protocol extensions of BGP (MP-BGP)~\cite{mpbgp-rfc4760}. 
	PEs run MP-BGP between each other, and use it for encoding and distributing the mappings in order to populate VRF tables.  
	
		\vspace{1mm}\noindent \textbf{2.1.~Associating interfaces to VPNs}: This task consists in tagging similarly interfaces that are part of the same VPN. In BGP MPLS VPNs, each interface is associated by configuration to a VRF, and each VRF is associated by configuration to a VPN customer. This is done by tagging, by configuration, each VRF with one or more numerical tags, called the Route Targets (RT). Two VRFs that are part of the same VPN are tagged with the same Route Targets\footnote{In practice, two different types of Route targets are needed to implement complex VPN topologies (e.g. hub and spoke).}.
		
		\vspace{1mm}\noindent \textbf{2.2.~Mappings Distribution}: The mapping (Site IP, PE address or label, interface label) relative to a VPN interface is encoded in a BGP route and is flooded to populate distant VRFs of the same VPN. The PE address is stored in the BGP Next Hop attribute. The site IP together with the internal label are encoded in the BGP NLRI field. 
		Finally, the Route Target (RT) of the VPN interface is appended to the BGP route (encoded as an extended BGP community). This BGP route is then flooded (distributed to all the PEs in the provider backbone). However, only distant VRFs that are tagged with the same Route Target will ultimately install it. 
	
\vspace{1mm}\noindent \textbf{3.~Connecting an ingress interface with a specific egress interface}: This is done using MPLS thanks to the stacking of two MPLS labels. The traffic that comes from one ingress interface is encapsulated in an MPLS tunnel with two labels that are pushed. The outer label allows the traffic to be routed through the provider network till the egress PE router, whereas the inner label allows the egress PE to identify to which egress interface the traffic should be forwarded. This explains why the interface locator is defined by two labels: one global egress PE identifier, and one egress interface label that is local to the PE. 

\vspace{1mm}\noindent \textbf{4.~Learning site ids behind provider interfaces}: In BGP MPLS VPNs, this is done thanks to the PE-CE routing protocol that runs between the VPN site CE router and the provider network PE router that connects it (can be a link state protocol like ISIS or OSPF, or a protocol like BGP, or even static routing).

\subsubsection{The virtual router approach}
\begin{figure}[tb]
	\centering
		\includegraphics[scale=0.25]{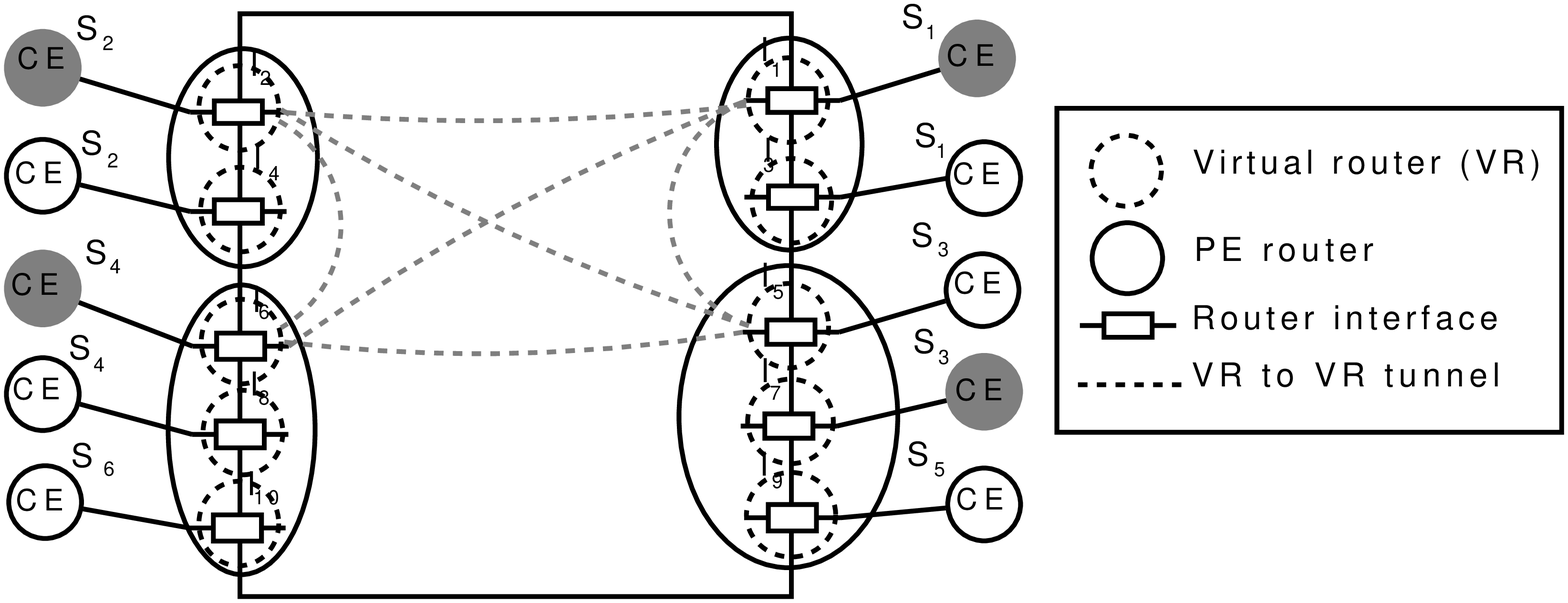}
		\caption{The virtual router approach (VR-to-VR model)}\label{fig:vr1}
\end{figure}

The virtual router approach~\cite{VR:draft} is another proposal that offers the same IP VPN service. Fig.~\ref{fig:vr1} illustrates a Virtual router implementation of the VPN service on top of our simple provider network model. We show how it is also centered around the elements that we depicted in the generic solution of Sec.~\ref{solution}.

\vspace{1mm}\noindent \textbf{1.~A per interface routing table}. The virtual router approach associates a Virtual Router (VR) per interface. A \textit{virtual router} is an emulation of a physical router. It has therefore its same capabilities including a routing table.
	
	\vspace{1mm}\noindent \textbf{2.~A method to populate per interface routing tables}: In the virtual routing approach, the CE routers together with the VRs of the same VPN form a Virtual Network. As such, They can run whatever routing protocol to populate the VR (and therefore the per-interface) routing tables.
	
		\vspace{1mm}\noindent \textbf{2.1.~Associating interfaces to VPNs}: In the VR approach, tagging similarly interfaces that are part of the same VPN results in connecting the VRs together to form the VPN topology. The VR approach proposes to use an auto discovery mechanism~\cite{vpn-bgp} (with Route Targets similarly to BGP MPLS VPNs), to allow each virtual router to discover which distant VRs belong to the same VPN. In short, VRs that are part of the same VPN are configured with Route Targets. BGP messages containing VR identifiers (interface locators) and their Route Targets are flooded/exchanged in the provider network so that each VR discovers all the other VRs that are part of its VPN.
		 
		\vspace{1mm}\noindent \textbf{2.2.~Mappings Distribution}: Once VRs discover which distant VRs belong to the same VPN, they use one of the tunneling techniques (in point 3. below) to exchange both routing information and data. VRs of the same VPN form a virtual network and can run any routing protocol between each other to build the mappings between site ids and interface locators (These mappings are the routing entries of the routing table of each VR). 
	
	\vspace{1mm}\noindent \textbf{3.~Connecting an ingress interface with a specific egress interface}: Connecting two interfaces is done with tunneling like with BGP MPLS VPNs. However, the VR approach did not impose a tunneling technique, which can be IPSec~\cite{ipsec}, IP-in-IP~\cite{ipinip}, GRE~\cite{gre}, MPLS~\cite{MPLS_rfc}, or even a layer-2 connections (e.g. ATM). 
	
	\vspace{1mm}\noindent \textbf{4.~Learning site ids behind provider interfaces}: The routing protocol that runs between the VR and the CE allows the VR to learn site ids.

\subsection{The case of Layer-2 VPNs}
In layer-2 VPNs~\cite{rfc4664}, similarly to layer 3 VPNs, the provider backbone has a set of PE routers. Each PE has a set of interfaces that connect customer sites. 
There are mainly two types of layer-2 VPNs, virtual private wire service (VPWS) and virtual private LAN service (VPLS).
With VPWS, a provider backbone offers a point-to-point layer-2 connection (e.g to transport ATM or Ethernet) between a pair of sites that need to be interconnected. VPLS is instead a point-to-multipoint layer-2 Ethernet VPN service that emulates a LAN.
They both exhibit the same pattern as in the generic solution. However, due to lack of space, we consider only the VPLS approach.

\subsubsection{Virtual Private LAN Service (VPLS)}
\begin{figure}[h]
	\centering
		\includegraphics[scale=0.25]{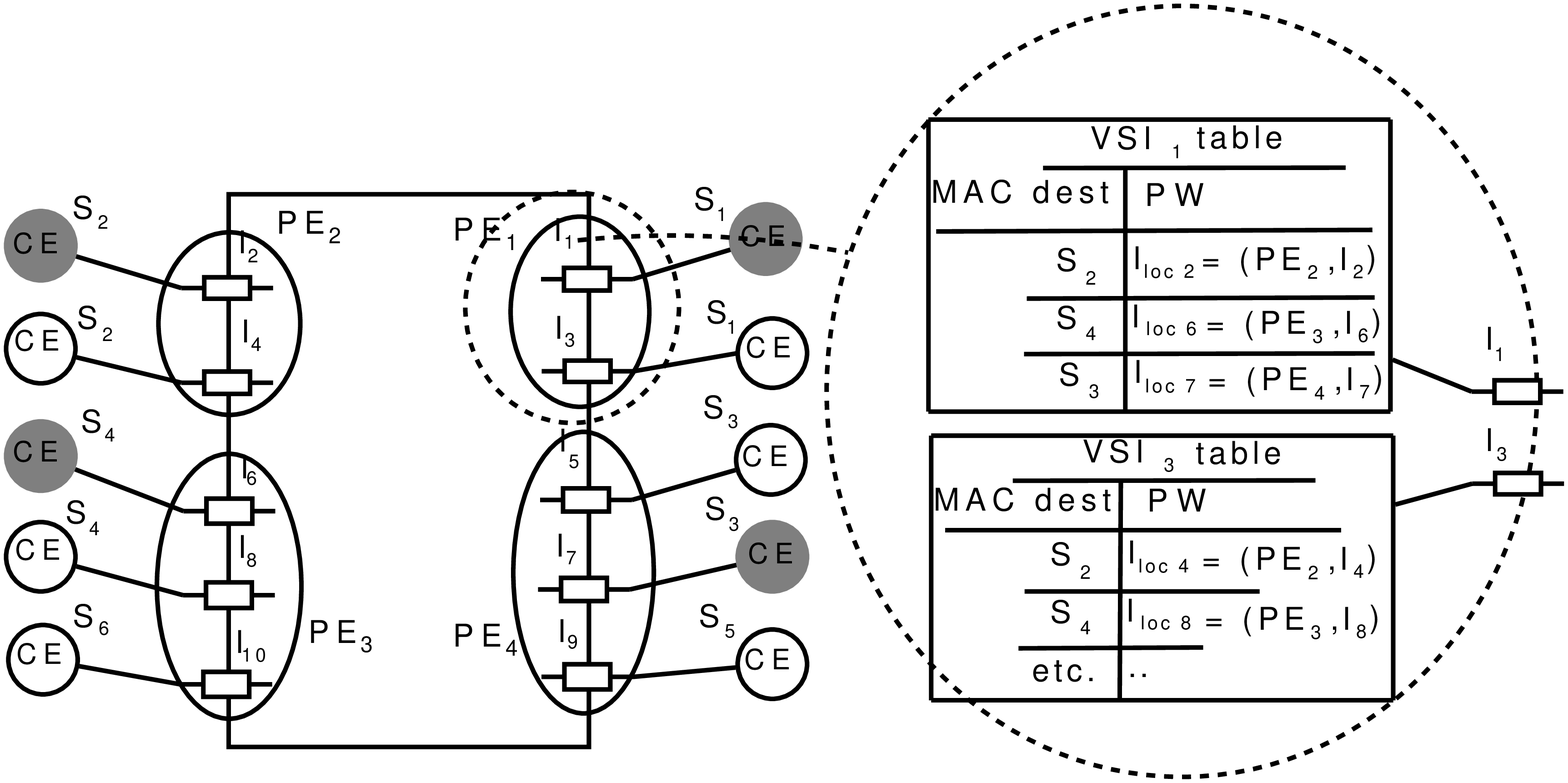}
		\caption{VPLS implementation}\label{fig:vpls}
\end{figure}

Fig.~\ref{fig:vpls} shows a VPLS implementation of the VPN service on the top of the provider network described in Fig.~\ref{fig:interface_matrix}. With respect to our model, site ids are the set of MAC addresses of the site. We show that also VPLS comprise the four elements depicted in our generic solution.

	\vspace{1mm}\noindent \textbf{1.~A per interface routing table}. In VPLS, each interface has a Virtual Switching Instance (VSI), a table that contains mappings between site ids (destination MAC addresses) and the corresponding interface locators (PWs) to which the traffic should be forwarded.
	
	\vspace{1mm}\noindent \textbf{2.~A method to populate per interface routing tables}: In VPLS also, this is done in the two steps that we depicted earlier.
	
			\vspace{1mm}\noindent \textbf{2.1.~Associating interfaces to VPNs}: tagging similarly interfaces that are part of the same VPN is done using either LDP~\cite{rfc4762} or BGP~\cite{rfc4762} (two dedicated RFCs were issued for the VPLS case). If BGP is used, this is done exactly similarly to the VR approach in layer 3 VPNs. The result of this step is that each ingress interface is associated with a list of egress interfaces to which it is connected.
			
	
		\vspace{1mm}\noindent \textbf{2.2.~Mappings Distribution}: In the classical VPLS standard, the mappings are not explicitly distributed as with BGP MPLS VPNs. Instead, each VSI acts as a layer 2 switch and performs MAC learning to associate site ids (MAC addresses) to the corresponding interface locators (PWs).
			In theory, the mappings distribution could be done explicitly, exactly the same way as with BGP MPLS VPNs, by encoding (site ids, interface locators) in BGP routes. This is the purpose of a very recent solution that is currently being developed~\cite{MAC-VPN} in the IETF.
		
	\vspace{1mm}\noindent \textbf{3.~Connecting an ingress interface with a specific egress interface}: This is done similarly to BGP MPLS VPNs with a stacking of two MPLS labels.
	
	
	\vspace{1mm}\noindent \textbf{4.~Learning site ids behind provider interfaces}: VSIs perform MAC learning to learn site ids (MAC addresses) of the VPN site.


\subsection{The case of Layer-1 VPNs}
The goal of Layer-1 VPNs is to provide point-to-point transport connections between pairs of distant VPN sites. 
The advent of such a solution is a consequence of the automation of layer 1 connections set up. Indeed, GMPLS~\cite{gmpls} proposes to use an IP control plane to remotely control GMPLS capable transport devices and instruct them to fast establish layer-1 connections.   
Work on layer-1 VPNs was carried jointly within both the ITU-T and the IETF. The ITU-T first standardized layer-1 VPN requirements and high level architecture~\cite{l1vpn:itu2,l1vpn:itu} and the IETF worked later on protocol aspects~\cite{rfc5251,bgp:l1}. We now show how also layer 1 VPNs exhibit the same pattern as our generic solution. In layer-1 VPNs, customer sites have ports that connect to the provider backbone through provider ports. The customer port has a Customer Port Identifier (CPI) that is the equivalent of the site id in our generic model. The provider port is equivalent to the provider VPN interface in our model. It can be identified in a unique way within the provider backbone. Each provider port has a provider port identifier (PPI), it is the equivalent to the interface locator in our model.

	\vspace{1mm}\noindent \textbf{1.~A per interface routing table}. In layer-1 VPN, a PE stores a port information table (PIT) per VPN port (equivalent to the VRF in BGP/MPLS IP VPNs and to the VSI in layer-2 VPLS). It contains (CPI,PPI) mappings.
	
	\vspace{1mm}\noindent \textbf{2.~A method to populate per interface routing tables}: This is done very similarly to BGP MPLS VPNs using extensions to BGP for layer 1 VPNs~\cite{bgp:l1}.

		\vspace{1mm}\noindent \textbf{2.1.~Associating interfaces to VPNs}: tagging similarly interfaces that are part of the same VPN is done using Route Targets as with BGP MPLS VPNs.

		\vspace{1mm}\noindent \textbf{2.2.~Mappings Distribution}: (CPI,PPI) mappings are encoded in BGP routes together with Route Target information exactly as done with BGP MPLS VPNs.

	\vspace{1mm}\noindent \textbf{3.~Connecting an ingress interface with a specific egress interface}: Thanks to GMPLS, it is possible, through IP based signaling mechanisms, to automatically build a layer-1 physical link between any two given provider ports. 

	\vspace{1mm}\noindent \textbf{4.~Learning site ids behind provider interfaces}: This is done thanks to the GMPLS IP Control Channel (IPCC) between the provider equipment and the customer equipment.
\section{Towards knowledge reuse and automated reasoning}\label{overview}
We believe that networking should become more than struggling with complex packet header formats, hundred-page cumbersome specification documents with inconsistent terminologies, or endless religious debates about whether ICN or SDN are the future of the discipline. The previous section was a modest attempt to show that there is a lack of knowledge reuse in our discipline, and that we are may be solving the same problems over and over again. We see two possible approaches to effectively start moving the opposite way. The first is theoretical, the second is a practical approach to tackle the first.

\subsection{A theory of networking}\label{theory}
The first approach is to build a ``theory of networking'', not a theory that captures one aspect of networking (like routing protocols or congestion control), but something that puts together (as much as possible) all the small pieces (e.g addressing, naming, DNS, routing, bridging, tunneling, policies, management, reaction to failures etc.). Such a theory would capture the big picture, and position networking with respect to neighboring disciplines. This should define what is within the scope of networking and what is not. 

In order to do so, we need to first progress our understanding of the axioms of networking today: what are its assumptions? and what problems does networking aim to solve under these assumptions? Once we understand them and agree on them, we need a formalism to represent them and to represent the networking problems. The next step would be then to resolve all old networking problems in the light of this new formalism. If such an ``axiomatisation'' is well done, it would be possible to ``reinvent'' what we know about networking under this formalism. A positive side effect of this approach would be to discover classes of networking problems and solutions that can be more easily reused. Since any networking problem can be formulated under this framework, the reuse of previous solutions and proofs should be easier.

To give a concrete example, one reasonable first axiom for networking could be to assume that two hosts equipped with computing, storage and networking capabilities (an interface) are able to transfer data between their respective memory descriptors (read bytes from one memory area and send to another memory area) if they are instructed by a user to do so. This axiom would separate networking from other disciplines like signal, coding theory etc.  Another axiom would define the naming of these memory descriptors, their scope, and which entity defines them so as to realize a communication between two hosts. Networking would be then the science of making these two hosts communicate, and generalizing this case from two to multiple hosts.

Nevertheless, this first approach is a bit arduous and may be too ambitious. It is somehow doing to networking what Principia Mathematica~\cite{principia} attempted to do with mathematics: ``reinventing'' it from scratch starting from the smallest set of axioms. We next suggest a more practical ``trial and error'' way to approach this problem.

\subsection{An automated networking expert}\label{expert_system}
A practical way to approach the more abstract problem described above is to leverage another discipline: automated reasoning, a sub field of Artificial intelligence. Although Artificial intelligence failed to mechanize human thinking and reasoning at large (Strong AI or artificial general intelligence), it fairly succeeded to automate specific aspects of reasoning (e.g. reasoning about a specific area or type of problems). This has been the case for automated theorem proving~\cite{automated:theorem} in mathematics, but also for expert systems which proved their efficiency in reasoning about domain specific problems. 

We could learn a lot by playing with implementing expert systems (or any semantic reasoner) that reason about networking problems, starting from small basic problems, and increasing the complexity. 
Similarly to traditional expert systems, such a reasoner would be fed with a \textit{knowledge database} that contains the assumptions we have about networking (or the ``facts''). Next, an \textit{inference} or \textit{rules engine} takes as an input the facts in the knowledge database to produce new ones and so forth. Iterating this approach systematically by changing the facts in the knowledge database and the rules in the inference engine could help with understanding what problems can be solved with which set of assumptions and solutions.



\subsection{Risk and reward}\label{reward}
To conclude, what we advocate for in this paper is surely ambitious and difficult. However, we believe that it is worth trying. Even if we don't reach the goal of building an automated networking reasoner, we could answer, on the way, questions that will enhance our understanding of our discipline. For example, why did ``mankind'' invent IP addresses, VLANs, MAC, MPLS, and different combinations of encapsulations like MAC in MAC, MAC in MPLS, IP in MPLS? Was it fundamentally necessary to have so many networking identifiers? Or was this simply a historical heritage? What is the minimal set of fundamentally necessary identifiers/names to provide each networking function? MPLS, for instance, was initially invented to speed up packet switching since IP lookup was costly~\cite{huston-mpls}. Due to the progress of technology, this reason vanished, but MPLS found a life in satisfying other networking functions, like fine grained traffic engineering, and VPN routing. But if the primary goal was to provide the VPN service or to do traffic engineering, would MPLS have been necessary? What if we had just used IP tunneling instead of MPLS?

Another useful outcome is to understand whether it's possible and what would happen if we systematically relax each of the axioms of networking today. Indeed, relaxing an axiom could open the way for interesting ``discoveries''\footnote{somehow like relaxing Euclid's 5$^{th}$ postulate gave birth to interesting non-euclidean geometries}. 
This is by the way what Content Centric Networking (CCN) did to the discipline. Indeed, it relaxed one of the axioms of traditional networking, namely the fact that a name to host resolution step precedes a host to host communication. CCN not only showed, by its feasibility, that the ``name to address'' resolution step is not a must for networking to ``work'', but also opened the way for a parallel networking discipline, with its own new transport protocols, routing protocols, routers and naming schemes to name few.

Finally, such an approach can have positive effects on the way this discipline is taught, by allowing students to see the big picture, where the different pieces fit, and how they interact together, instead of getting lost in the nasty details of specific protocol details. 
\bibliography{report-bib}
\bibliographystyle{ieeetr}
\end{document}